\documentclass [12pt,a4paper]{article}
\usepackage{amssymb, theorem}
\def\bbbc{{\mathchoice {\setbox0=\hbox{$\displaystyle\rm C$}\hbox{\hbox
to0pt{\kern0.4\wd0\vrule height0.9\ht0\hss}\box0}}
{\setbox0=\hbox{$\textstyle\rm C$}\hbox{\hbox
to0pt{\kern0.4\wd0\vrule height0.9\ht0\hss}\box0}}
{\setbox0=\hbox{$\scriptstyle\rm C$}\hbox{\hbox
to0pt{\kern0.4\wd0\vrule height0.9\ht0\hss}\box0}}
{\setbox0=\hbox{$\scriptscriptstyle\rm C$}\hbox{\hbox
to0pt{\kern0.4\wd0\vrule height0.9\ht0\hss}\box0}}}}

\def\qed{{\rightline \sq}\medskip}

\newtheorem{prop}{Proposition}

\theorembodyfont{\rm}

\renewcommand{\qed}{\nobreak\hfill $\square$\medskip}
\def\proof{\noindent{\it Proof.} }
\def\<{\langle}
\def\>{\rangle}

\def\iH{{\cal H}}
\def\iM{{\cal M}}

\def\iA{{\cal A}}

\def\b1{{\bf 1}}

\def\Re{{\rm Re}\,}
\def\Im{{\rm Im}\,}
\def\Tr{{\rm Tr}}

\def\im{{\rm i}}
\def\ot{\otimes}

\def\Tr{\mbox{Tr}\,}

\begin{document}
\vspace{-2cm}
\ \vskip 1cm
\centerline{\LARGE {\bf State tomography for two qubits}}
\bigskip
\centerline{\LARGE {\bf using reduced densities}}
\bigskip
\medskip\footnotetext[3]{Supported by the Hungarian Research Grants OTKA 
T042710, T063066 and T032662.}
\bigskip
\centerline{ D. Petz$^{1,3}$, K.M. Hangos$^{2,3}$, A. Sz\'ant\'o$^{1,3}$ and 
F. Sz\"oll\H osi$^{1,3}$}
\medskip
\bigskip
\centerline{$^1$ Budapest University of Technology and Economics}
\centerline{$^2$ Computer and Automation Research Institute}
\medskip
\bigskip
\begin{quote}
{\bf Abstract:} The optimal state determination (or tomography) is 
studied for a 
composite system of two qubits when measurements can be performed 
on one of the qubits and interactions of the two qubits can be 
implemented. The goal is to minimize the number of interactions to
be implemented. The algebraic method used in the paper leads to an 
extension of the concept of mutually unbiased measurements.



\end{quote}
\bigskip
\section{Introduction}
An $n$-level quantum system is described by an $n$-dimensional Hilbert
space $\iH$, or equivalently by the algebra $M_n(\bbbc)$ of the $n \times n$
complex matrices. When an orthogonal basis of $\iH$ is chosen, operators
acting on $\iH$ correspond to $n \times n$ matrices. A positive operator 
$\rho$ of trace 1 is called state. If we choose and fix an orthonormal 
basis $\{e_1,e_2,\dots, e_n\}$, then a state $\rho$ is determined by the 
matrix elements $\rho_{ij}= \< e_i|\rho|e_j\>$. Determination of $\rho$ 
involves $n^2-1$ real parameters, namely, $\rho_{ii}$ ($ 1 \le i \le n-1$), 
$\Re \rho_{ij}$ and $\Im \rho_{ij}$ ($1 \le i < j \le n$).

A von Neumann measurement on the system is a family $\iM=\{P_1, P_2,
\dots,P_d\}$ of pairwise orthogonal projections such that $\sum_i P_i =I$.
When the measurement $\iM$ is performed in the state $\rho$, the outcome
$1 \le j \le d$ appear with probability $p_j=\Tr \rho P_i$ for each 
$j$ \cite{N-Ch, OP}. 
Independent measurements on several copies of our quantum system give
the relative frequencies $f_j$ for each outcome $j$ and $f_j$ is an estimate 
of the probability $p_j$. The repeated measurement provides $d-1$ degree
of freedom concerning the the density $\rho$, since $\sum_j p_j=1$. The
information we obtained is maximal if $d=n$ which means that all the
projections $P_j$ are of rank one. $\rho$ is determined by $n^2-1$
parameters, hence at least $n+1$ different measurements are to be performed 
to cover all degrees of freedom. Of course,  the $n+1$  different 
measurements are sufficient in the case when they provide ``non-overlapping'' 
information. 

The literature of state tomography is very rich, there are
several protocols, and the efficiency of state reconstruction can be increased
if the later measurements depend on the outcomes of the former ones \cite{
Buzek, book,REK}.

The composite system of two qubits is a 4-level quantum system
which is described on the space $\bbbc^4 \equiv \bbbc^2 \otimes \bbbc^2$.
A state is described by 15 real parameters. Therefore, at least 5 kinds of 
elementary measurements should be made to determine the state of the system.

Denote by $A$ and $B$ the two qubits. Then $M_4(\bbbc)=B(\iH_A)\otimes
B(\iH_B)$, where $B(\iH_A)$ and $B(\iH_B)$ are isomorphic to $M_2(\bbbc)$.
Assume that we can perform measurements only on the qubit $A$. If the total 
system has the statistical operator $\rho_{AB}$, then we can reconstruct 
the reduced density 
$$
\rho_A^{(1)}:=\Tr_B \rho_{AB}
$$ 
after some measurements. In order to get more information, we switch 
on an interaction between the two qubits. If $H$ is the Hamiltonian, then 
the new state is
\begin{equation}\label{1}
e^{\im H} \rho_{AB} e^{-\im H}= W_1 \rho_{AB} W_1^*
\end{equation}
after the interaction. (For the sake of simplicity, the interaction is kept
for a time unit.) The new reduced density is
$$
\rho_A^{(2)}:=\Tr_B W_1 \rho_{AB}W_1^*.
$$
This procedure may be continued by using other interactions and ends with a 
sequence of reduced states $\rho_A^{(1)},\allowbreak \rho_A^{(2)},\dots, 
\rho_A^{(k)}$. We want to 
determine the minimal $k$ such that this sequence of reduced densities 
determines  $\rho_{AB}$. In other words, we want to minimize the number of
interactions between the two qubits. It turns out that the minimum number
is 5. 

Minimal realizations play an important role in systems theory, too \cite{Isi},
 because they represent the state of the system with the minimum
 possible number of parameters.
Minimal realizations are known to be jointly controllable
 and observable for most of the known system classes.
The above problem of finding the minimum number of reduced
 states can be regarded as a minimal representation
 problem for a system that consists of a pair of coupled qubits.

\section{Algebraic formulation}

Instead of the transformation (\ref{1}) of the density matrix $\rho_{AB}$, 
we can change the subalgebra and we have an equivalent algebraic formulation.
The total system is described by the algebra $M_4(\bbbc)$. We look for 
subalgebra $\iA_1, \iA_2, \dots, \iA_k$ such that
\begin{enumerate}
\item
Each $\iA_j$ is algebraically isomorphic to $M_2(\bbbc)$, $1 \le j \le k$.
\item
The  linear span of the subspaces $\iA_1, \iA_2, \dots, \iA_k$ is  
$M_4(\bbbc)$.
\end{enumerate}
Given a subalgebra $\iA_j$, there is a unitary $W_j$ such that $W_j^*\iA_jW_j=
B(\iH_A)\otimes \bbbc I_B$. The reduced density $\rho_j \in \iA_j$ is
the same as the reduction of $W_j\rho_{AB} W_j^*$ to the first spin. Therefore,
instead of the reduction of density after the interaction, we can work with the
reduced density of $\rho_{AB}$ in $\iA_j$. The second condition makes sure 
that the reduced densities in $\iA_1, \iA_2, \dots, \iA_k$ determine 
$\rho_{AB}$ completely.

The traceless subspace of $M_4(\bbbc)$ has dimension 15, while the
traceless subspace of $\iA_j$ has dimension 3, therefore we need 
$k\ge 5$ to fulfill the requirements. It will turn out that $k=5$ is
possible.

The algebra $M_2(\bbbc)$ is linearly spanned by the Pauli matrices:
$$
\sigma_0 :=
\left[ \begin{array}{cc} 1&0\\ 0&1\end{array} \right], \quad
\sigma_1 :=
\left[ \begin{array}{cc} 0&1\\ 1&0\end{array} \right], \quad
\sigma_2 :=
\left[ \begin{array}{cc} 0&-\im\\ \im&0\end{array} \right], \quad 
\sigma_3:=
\left[ \begin{array}{cc} 1&0\\ 0&-1\end{array} \right].
$$
Recall that they satisfy the multiplication rules
\begin{equation}\label{E:Levi}
\sigma_i \sigma_j=\delta_{ij}I+\im \sum_{k=1}^3 \epsilon_{ijk}\sigma_k
\qquad (1 \le i,j \le 3),
\end{equation}
where $\epsilon_{ijk}$ is the Levi-Civita tensor:
$$
\epsilon_{i_1 i_2 \cdots i_n}=\left\{
\begin{array}{ll}
0 & \textrm{if $\exists j,k$ such that $i_j=i_k$,}\\
1 & \textrm{if the permutation $(i_1 i_2 \cdots i_n)$ is even,} \\
-1 & \textrm{if the permutation $(i_1 i_2 \cdots i_n)$ is odd.}\\
\end{array}
\right.
$$
The above rules can be essentially simplified by posing the following two
requirements:
\begin{enumerate}
\item
$\sigma_j$ is a self-adjoint unitary ($1 \le j \le k$) and
$\sigma_3= - \im\sigma_1 \sigma_2$.
\item
$\sigma_1 \sigma_2 +\sigma_2 \sigma_1=0$.
\end{enumerate}
When a triplet $(S_1,S_2,S_3)$ satisfies these condition, it will be called 
a Pauli triplet. For such a triplet $\Tr S_i=0$ and $\Tr S_iS_j=0$ for 
$i \ne j$. The latter relation is interpreted as the orthogonality of
$S_i$ and $S_j$ with respect to the Hilbert-Schmidt inner product $\< A,B\>:
= \Tr A^*B$. Furthermore, it can be seen that the two relations above imply
(\ref{E:Levi}).

Given a Pauli triplet $(S_1,S_2,S_3)$, the linear mapping
defined as
$$
\sigma_0 \mapsto I,\quad \sigma_1 \mapsto S_1, \quad 
\sigma_2 \mapsto S_2, \quad \sigma_3 \mapsto -\im S_1S_2
$$
is an algebraic isomorphism between $M_2(\bbbc)$ and the linear span of
the operators $I,S_1,S_2$ and $S_3$.

In the algebra $M_4(\bbbc)$, the elementary tensors $\sigma_i \otimes \sigma_j$
form an orthogonal basis ($0 \le i,j \le 3$). All these operators
are self-adjoint unitaries and can be chosen to be $S_i$'s.

The next proposition is the main result of the paper.

\begin{prop}\label{prop1}
There are 5 sublalgebras of $B(\iH_A)\ot B(\iH_B)$ such that each of
them is isomorphic to $M_2(\bbbc)$ and the reduced states determine an 
arbitrary state $\rho_{AB}$ of the two qubits $A$ and $B$.
\end{prop}

{\it Proof.}
First we take the following Pauli triplets consisting of elementary
tensors:
\begin{eqnarray*} \label{felb}
&& \{ \sigma_0\otimes\sigma_1,-\sigma_1\otimes\sigma_3,
\sigma_1\otimes\sigma_2\} \cr && \phantom{MMMMMMMM} \cr 
&& \qquad \qquad  =\left\{\left[\begin{array}{rrrr}
0 & 1 & 0 & 0\\
1 & 0 & 0 & 0\\
0 & 0 & 0 & 1\\
0 & 0 & 1 & 0\\
\end{array}\right],
\left[\begin{array}{rrrr}
0 & 0 & -1 & 0\\
0 & 0 & 0 & 1\\
-1 & 0 & 0 & 0\\
0 & 1 & 0 & 0\\
\end{array}\right],
\left[\begin{array}{rrrr}
0 & 0 & 0 & -\im\\
0 & 0 & \im & 0\\
0 & -\im & 0 & 0\\
\im & 0 & 0 & 0\\
\end{array}\right]\right\}, \cr && \phantom{MMMMMMMM} \cr
&& \{\sigma_3\otimes\sigma_1,\,\sigma_1\otimes\sigma_1,\,
\sigma_2\otimes\sigma_0\} \cr && \phantom{MMMMMMMM}  \cr && \qquad \qquad
= \left\{\left[\begin{array}{rrrr}
0 & 1 & 0 & 0\\
1 & 0 & 0 & 0\\
0 & 0 & 0 & -1\\
0 & 0 & -1 & 0\\
\end{array}\right],
\left[\begin{array}{rrrr}
0 & 0 & 0 & 1\\
0 & 0 & 1 & 0\\
0 & 1 & 0 & 0\\
1 & 0 & 0 & 0\\
\end{array}\right],
\left[\begin{array}{rrrr}
0 & 0 & -\im & 0\\
0 & 0 & 0 & -\im\\
\im & 0 & 0 & 0\\
0 & \im & 0 & 0\\
\end{array}\right]
\right\}, \cr && \phantom{MMMMMMMM} 
\cr && 
\{\sigma_1\otimes\sigma_0,\,\sigma_2\otimes\sigma_2,\,
\sigma_3\otimes\sigma_2\} \cr && \phantom{MMMMMMMM}   \cr && \qquad\qquad
=\left\{\left[\begin{array}{rrrr}
0 & 0 & 1 & 0\\
0 & 0 & 0 & 1\\
1 & 0 & 0 & 0\\
0 & 1 & 0 & 0\\
\end{array}\right],
\left[\begin{array}{rrrr}
0 & 0 & 0 & -1\\
0 & 0 & 1 & 0\\
0 & 1 & 0 & 0\\
-1 & 0 & 0 & 0\\
\end{array}\right],
\left[\begin{array}{rrrr}
0 & -\im & 0 & 0\\
\im & 0 & 0 & 0\\
0 & 0 & 0 & \im\\
0 & 0 & -\im & 0\\
\end{array}\right]
\right\},\cr && \phantom{MMMMMMMM}
\cr && 
\{\sigma_0\otimes\sigma_2,\,\sigma_2\otimes\sigma_3,\,
\sigma_2\otimes\sigma_1\}  \cr && \phantom{MMMMMMMM}   \cr && \qquad \qquad
=\left\{\left[\begin{array}{rrrr}
0 & -\im & 0 & 0\\
\im & 0 & 0 & 0\\
0 & 0 & 0 & -\im\\
0 & 0 & \im & 0\\
\end{array}\right],
\left[\begin{array}{rrrr}
0 & 0 & -\im & 0\\
0 & 0 & 0 & \im\\
\im & 0 & 0 & 0\\
0 & -\im & 0 & 0\\
\end{array}\right],
\left[\begin{array}{rrrr}
0 & 0 & 0 & -\im\\
0 & 0 & -\im & 0\\
0 & \im & 0 & 0\\
\im & 0 & 0 & 0\\
\end{array}\right]
\right\}.
\end{eqnarray*}
Together with the identity, each triplet linearly spans a subalgebra $\iA_j$ 
($1 \le j \le 4$). It is important to observe that all the matrices have 
vanishing diagonal, moreover the matrices are pairwise orthogonal, therefore
they are linearly independent.

If we find another Pauli triplet $(S_1,S_2,S_3)$ 
such that the diagonals are linearly 
independent, then we have a fifth algebra $\iA_5$ such that $\{\iA_k:
1 \le k \le 5\}$ spans linearly $M_4(\bbbc)$. Indeed, if $A$ is any matrix,
then we can find $T \in \iA_5$ such that $A-T$ has 0 diagonal and
this is in the linear hull of $\{\iA_j: 1 \le j \le 4\}$. It follows that
the reduced densities in $\{\iA_j: 1 \le j \le 5\}$ determines $\rho_{AB}$ 
uniquely. 

Here is an example of the above described triplet:
$$
\frac12\left[\begin{array}{rrrr}
1 & 1 & 1 & 1\\
1 & 1 & -1 & -1\\
1 & -1 & -1 & 1\\
1 & -1 & 1 & -1\\
\end{array}\right],\quad \frac12
\left[\begin{array}{rrrr}
1 & \im & \im & -1\\
-\im & -1 & -1 & \im \\
-\im & -1 & 1 & -\im\\
-1 & -\im & \im & -1\\
\end{array}\right],\quad \frac12
\left[\begin{array}{rrrr}
-1 & \im & 1 & \im\\
-\im & 1 & \im & 1\\
1 & -\im & 1 & \im\\
-\im & 1 & -\im & -1\\
\end{array}\right].
$$
These matrices are not elementary tensors (but they are Hadamard matrices
\cite{Zych} up to a constant multiple and were found by means of an 
exhaustive search algorithm on a computer). \qed

\begin{prop}
If all the matrices of the Pauli triplet generating the subalgebras 
$\iA_j$ are of the form $\pm \sigma_k\otimes \sigma_l$ ($0 \le k,l \le 3$),
then we need at least 6 triplets to span $M_4(\bbbc)$. 
\end{prop}

{\it Proof.}
Assume that a Pauli triplet $(T_1,T_2,T_3)$ in $M_4(\bbbc)$ is such that
every element is of the form $\pm \sigma_i \otimes \sigma_j$ ($0 \le  i,j 
\le 3$).

If $T_1=\pm \sigma_i \otimes \sigma_j$ and $T_2=\pm\sigma_k \otimes 
\sigma_l$, then
\begin{eqnarray*}
\pm \im\,T_3 =&&-\sum_{m,n}(\epsilon_{ikm}\epsilon_{jln} 
\sigma_m\otimes \sigma_n)\\ &&+
\im\Big(\delta_{ik}\sum_n(\epsilon_{jln} \sigma_0\otimes
\sigma_n)+\delta_{jl}\sum_m(\epsilon_{ikm} \sigma_m\otimes
\sigma_0)\Big)+\delta_{ik}\delta_{jl}\sigma_0\otimes \sigma_0.
\end{eqnarray*}
Since $T_3$ is self-adjoint but  $\im \sigma_i \otimes \sigma_j$
is not, it follows that exactly one of the relations $i=k$ and $j=l$ 
must hold. At least one of the operators $T_i$ should be of the form
$\sigma_0 \otimes \sigma_j$ or $\sigma_j \otimes \sigma_0$.

We have three operators in the form $\sigma_0 \otimes \sigma_j$ and
three in the form $\sigma_j \otimes \sigma_0$ ($1 \le j \le 3$). 
If we have 5 Pauli triplets, then at least one should contain two of 
the above tensor products (up to a sign). One can see that  
$\sigma_0 \otimes \sigma_j$ and $\sigma_i \otimes \sigma_0$ cannot be 
in a triplet, therefore a triplet contains two operators in the form 
$\sigma_0 \otimes \sigma_j$ or two operators like $\sigma_j \otimes \sigma_0$. 
In both cases, the third operator has similar form. Hence one of the 
operators $\sigma_0 \otimes \sigma_j$ and $\sigma_j \otimes \sigma_0$ appears 
in two triplets and in this case 5 triplet cannot span the whole space. 

Six subalgebras described in the proposition can be given by the following 
Pauli triplets:
\begin{equation} \label{felb2}
\begin{array}{c}
\{\sigma_1\otimes \sigma_1, \sigma_1\otimes \sigma_2, \sigma_0 \otimes
\sigma_3\},\\
\{\sigma_2\otimes \sigma_2, \sigma_2\otimes \sigma_3, \sigma_0 \otimes
\sigma_1\},\\
\{\sigma_3\otimes \sigma_3, \sigma_3\otimes \sigma_1,  \sigma_0 \otimes
\sigma_2\},\\
\{\sigma_2\otimes \sigma_2, \sigma_3\otimes \sigma_2,
\sigma_1\otimes  \sigma_0\},\\
\{\sigma_3\otimes \sigma_3, \sigma_1\otimes \sigma_3,
\sigma_2\otimes  \sigma_0\},\\
\{\sigma_1\otimes \sigma_1, \sigma_2\otimes \sigma_1,
\sigma_3\otimes  \sigma_0\}.\\
\end{array}
\end{equation}
Together with $I$ each triplet linearly spans a subalgebra $\iA_j$ 
($1 \le j \le 6$) and the 6 subalgebras linearly span the whole 
$M_2(\bbbc)\otimes M_2(\bbbc)$. \qed

\section{Generalizations}

Mutually unbiased bases (or measurements) are interesting from many point of
view \cite{Kr, BSTW, WF} and the maximal number of such bases is not completely
known \cite{We}. The above discussed setting of state determination is
somewhat similar. In this setting we may look for essentially orthogonal
non-commutativ subalgebras while unbiased elementary measurement are given
essentially orthogonal maximal Abelian subalgebras, see Prop. 2.2 of  
\cite{Parta}. The next statement is an analogue of Parthasarathy's 
proposition.

\begin{prop}\label{P:uj}
Let $\iA_1$ and $\iA_2$ be subalgebras of $M_n(\bbbc)$ and assume that
they are isomorphic to $M_k(\bbbc)$. Then the following conditions
are equivalent:
\begin{enumerate}
\item[(i)] 
If $P \in \iA_1$ and $Q \in \iA_2$ are minimal projections,
then $\Tr PQ=n/k^2$.
\item [(ii)]
The subspaces $\iA_1 \ominus \bbbc I$ and  $\iA_2 \ominus \bbbc I$
are orthogonal in $M_n(\bbbc)$
\end{enumerate}
\end{prop}

\proof
It follows from the conditions that $n=mk$ and $\Tr P=\Tr Q=m$ for
the minimal projections. Therefore, condition (i) is equivalent to
$\Tr (I-kP)(I-kQ)=0$ which means that $(I-kP)\perp (I-kQ)$. Since the
subspaces in (ii) are linearly spanned by these operators, the 
proposition follows. 
\qed

Now we generalize Prop. 2 for $n$ qubits.

\begin{prop}
If all the matrices of the Pauli triplet generating the subalgebras 
$\iA_j$ of $M_{2^n}(\bbbc)$ are of the form $\pm \sigma_{k(1)}\otimes \dots
\otimes \sigma_{k(n)}$ $(0 \le k(i) \le 3, 1\le i \le n)$,
then we need more than $(2^{2n}-1)/3$ triplets to span $M_{2^n}(\bbbc)$. 
\end{prop}

{\it Proof.}
First note that a matrix $\pm \sigma_{k(1)}\otimes \dots
\otimes \sigma_{k(n)}$ has only real elements or only imaginary elements.
Among the three matrices of a Pauli triplet $(T_1,T_2,T_3)$, there is one
imaginary or there are three imaginary matrices. Let $N$ be the number of
triplets with 1 imaginary matrix and $M$ be the number of triplets with
3 imaginary ones. If the $N+M$ triplets with identity linearly span
the self-adjoint subspace, then $3(N+M)+1\ge 2^{2n}$. Assume that 
\begin{equation}\label{e:1}
N+M=\frac{2^{2n}-1}{3}. 
\end{equation}
Since the dimension of the subspace of self-adjoint matrices with imaginary
elements is $(2^{2n}-2^n)/2$, we must have  
\begin{equation}\label{e:2}
N+3M=\frac{2^{2n}-2^n}{2}.
\end{equation}
One can see that equations (\ref{e:1}) and (\ref{e:2}) do not have integer
solution. \qed

We call a family $\iM_1, \iM_2,\dots, \iM_d$ of subalgebras strongly 
mutually unbiased if the conditions in the proposition hold for any pair. 
The maximal number of strongly mutually unbiased subalgebras is not know 
to us even in the simplest case when the large algebra is $M_4(\bbbc)$ and
the subalgebras are isomorphic to $M_2(\bbbc)$.

Following \cite{Parta}, we may call a family $\iM_1, \iM_2,\dots, \iM_d$ 
of subalgebras weakly mutually unbiased if the subspaces $\iM_1\ominus 
\bbbc I, \iM_2\ominus \bbbc I,\dots, \iM_d\ominus \bbbc I$ are linearly 
independent. We showed that when the large algebra is $M_4(\bbbc)$ and
the subalgebras are isomorphic to $M_2(\bbbc)$, then the maximum number
of weakly unbiased subalgebras is 5.

\section{Discussion and conclusions}

The optimal state tomography has been studied for a composite system of 
two qubits when measurements can be performed on one of the qubits and 
interactions of the two qubits can be implemented. Equivalently, we found
physically realizable minimal set of reduced densities. The transformation 
described by (\ref{1}) is realized by a properly designed measurement 
apparatus (see the experimental devices in \cite{REK}). Therefore, an 
additional, but still unsolved problem is to find the unitary to each of 
the reduced densities that transforms the state to be measured to another 
one belonging to the reduced density in question. (Some preliminary results 
on the effect of the real 4-dimensional rotation matrices on the reduced 
densities is found in \cite{SzA}.) 

The construction of 5 Pauli triplets of $4\times 4$ matrices from the tensor
products of Pauli matrices  contains heuristic steps combined with an 
exhaustive search for the missing final triplet.
This makes it practically impossible to generalize the method for higher 
dimensions. Even if the heuristic steps were removed the resulted algorithm
would fall to the NP-hard category because of the exhaustive search.

The questions we posed for two qubits can be asked about three (or
more qubits). The dimension of $M_8(\bbbc)\ominus \bbbc I$ is 63, so
we need at least 21 Pauli triplets to span the whole space. In the 
moment we can construct 22 spanning Pauli triplets by ad hoc method
(this will be discussed in another publication).

The problem investigated here motivates the definition of strongly and
weakly unbiased subalgebras. In relation with them, several questions
can be raised.

\end{document}